\documentclass[10pt,a4paper,onecolumn]{article}
\usepackage{marginnote}
\usepackage{graphicx}
\usepackage{xcolor}
\usepackage{authblk,etoolbox}
\usepackage{titlesec}
\usepackage{calc}
\usepackage{tikz}
\usepackage{hyperref}
\hypersetup{colorlinks,breaklinks,
            urlcolor=[rgb]{0.0, 0.5, 1.0},
            linkcolor=[rgb]{0.0, 0.5, 1.0}}
\usepackage{caption}
\usepackage{tcolorbox}
\usepackage{amssymb,amsmath}
\usepackage{ifxetex,ifluatex}
\usepackage{seqsplit}
\usepackage{xstring}

\usepackage{float}
\let\origfigure\figure
\let\endorigfigure\endfigure
\renewenvironment{figure}[1][2] {
    \expandafter\origfigure\expandafter[H]
} {
    \endorigfigure
}

\usepackage{fixltx2e} %
\usepackage[style=authoryear, backend=biber, hyperref=false]{biblatex}
\let\textscOrig=\textsc
\def\textsc#1{\expandafter\textscOrig{\seqsplit{#1}}}
\renewcommand{\seqinsert}{\ifmmode
  \allowbreak
  \else\penalty6000\hspace{0pt plus 0.02em}\fi}

\makeatletter
\let\href@Orig=\href
\def\href@Urllike#1#2{\href@Orig{#1}{\begingroup
    \def\Url@String{#2}\Url@FormatString
    \endgroup}}
\def\href@Notdoi#1#2{\def\tempa{#1}\def\tempb{#2}%
  \ifx\tempa\tempb\relax\href@Urllike{#1}{#2}\else
  \href@Orig{#1}{#2}\fi}
\def\href#1#2{%
  \IfBeginWith{#1}{https://doi.org}%
  {\href@Urllike{#1}{#2}}{\href@Notdoi{#1}{#2}}}
\makeatother

\newlength{\cslhangindent}
\newlength{\csllabelwidth}
 {%
  \setlength{\parindent}{0pt}
  \ifodd #1 \everypar{\setlength{\hangindent}{\cslhangindent}}\ignorespaces\fi
  \ifnum #2 > 0
  \setlength{\parskip}{#2\baselineskip}
  \fi
 }%
 {}
\usepackage{calc}

\usepackage[top=3.5cm, bottom=3cm, right=1.5cm, left=1.0cm,
            headheight=2.2cm, reversemp, includemp, marginparwidth=4.5cm]{geometry}

\titleformat{\section}
  {\normalfont\sffamily\Large\bfseries}
  {}{0pt}{}
\titleformat{\subsection}
  {\normalfont\sffamily\large\bfseries}
  {}{0pt}{}
\titleformat{\subsubsection}
  {\normalfont\sffamily\bfseries}
  {}{0pt}{}
\titleformat*{\paragraph}
  {\sffamily\normalsize}

\usepackage{fancyhdr}
\makeatletter
\let\ps@plain\ps@fancy
\fancyheadoffset[L]{4.5cm}
\fancyfootoffset[L]{4.5cm}

\definecolor{linky}{rgb}{0.0, 0.5, 1.0}

\newtcolorbox{repobox}
   {colback=red, colframe=red!75!black,
     boxrule=0.5pt, arc=2pt, left=6pt, right=6pt, top=3pt, bottom=3pt}

\newcommand{\ExternalLink}{%
   \tikz[x=1.2ex, y=1.2ex, baseline=-0.05ex]{%
       \begin{scope}[x=1ex, y=1ex]
           \clip (-0.1,-0.1)
               --++ (-0, 1.2)
               --++ (0.6, 0)
               --++ (0, -0.6)
               --++ (0.6, 0)
               --++ (0, -1);
           \path[draw,
               line width = 0.5,
               rounded corners=0.5]
               (0,0) rectangle (1,1);
       \end{scope}
       \path[draw, line width = 0.5] (0.5, 0.5)
           -- (1, 1);
       \path[draw, line width = 0.5] (0.6, 1)
           -- (1, 1) -- (1, 0.6);
       }
   }

\patchcmd{\@maketitle}{center}{flushleft}{}{}
\patchcmd{\@maketitle}{center}{flushleft}{}{}
\patchcmd{\@maketitle}{\LARGE}{\LARGE\sffamily}{}{}
\def\maketitle{{%
  
  \AB@maketitle}}
\makeatletter
\renewcommand\AB@affilsepx{ \protect\Affilfont}
\renewcommand\AB@affilnote[1]{{\bfseries #1}\hspace{3pt}}
\renewcommand{\affil}[2][]%
   {\newaffiltrue\let\AB@blk@and\AB@pand
      \if\relax#1\relax\def\AB@note{\AB@thenote}\else\def\AB@note{#1}%
        \setcounter{Maxaffil}{0}\fi
        \begingroup
        \let\href=\href@Orig
        \let\textsc=\textscOrig
        \let\protect\@unexpandable@protect
        \def\thanks{\protect\thanks}\def\footnote{\protect\footnote}%
        \@temptokena=\expandafter{\AB@authors}%
        {\def\\{\protect\\\protect\Affilfont}\xdef\AB@temp{#2}}%
         \xdef\AB@authors{\the\@temptokena\AB@las\AB@au@str
         \protect\\[\affilsep]\protect\Affilfont\AB@temp}%
         \gdef\AB@las{}\gdef\AB@au@str{}%
        {\def\\{, \ignorespaces}\xdef\AB@temp{#2}}%
        \@temptokena=\expandafter{\AB@affillist}%
        \xdef\AB@affillist{\the\@temptokena \AB@affilsep
          \AB@affilnote{\AB@note}\protect\Affilfont\AB@temp}%
      \endgroup
       \let\AB@affilsep\AB@affilsepx
}
\makeatother

\renewcommand\Affilfont{\sffamily\small\mdseries}
\ifnum 0\ifxetex 1\fi\ifluatex 1\fi=0 %
  \usepackage[T1]{fontenc}
  \usepackage[utf8]{inputenc}

\else %
  \ifxetex
    \usepackage{mathspec}
  \else
    \usepackage{fontspec}
  \fi
  \defaultfontfeatures{Ligatures=TeX,Scale=MatchLowercase}

\fi
\IfFileExists{upquote.sty}{\usepackage{upquote}}{}
\IfFileExists{microtype.sty}{%
\usepackage{microtype}
\UseMicrotypeSet[protrusion]{basicmath} %
}{}

\usepackage{hyperref}
\hypersetup{unicode=true,
            pdftitle={wolensing: A Python package to compute wave optics amplification factor for gravitational wave},
            pdfborder={0 0 0},
            breaklinks=true}
\let\addcontentslineOrig=\addcontentsline
\def\addcontentsline#1#2#3{\bgroup
  \let\textsc=\textscOrig\addcontentslineOrig{#1}{#2}{#3}\egroup}
\let\markbothOrig\markboth
\def\markboth#1#2{\bgroup
  \let\textsc=\textscOrig\markbothOrig{#1}{#2}\egroup}
\let\markrightOrig\markright
\def\markright#1{\bgroup
  \let\textsc=\textscOrig\markrightOrig{#1}\egroup}

\usepackage{graphicx,grffile}
\makeatletter
\def\maxwidth{\ifdim\Gin@nat@width>\linewidth\linewidth\else\Gin@nat@width\fi}
\def\maxheight{\ifdim\Gin@nat@height>\textheight\textheight\else\Gin@nat@height\fi}
\makeatother
\setkeys{Gin}{width=\maxwidth,height=\maxheight,keepaspectratio}
\IfFileExists{parskip.sty}{%
\usepackage{parskip}
}{%
\setlength{\parindent}{0pt}
\setlength{\parskip}{6pt plus 2pt minus 1pt}
}
\ifx\paragraph\undefined\else
\let\oldparagraph\paragraph
\renewcommand{\paragraph}[1]{\oldparagraph{#1}\mbox{}}
\fi
\ifx\subparagraph\undefined\else
\let\oldsubparagraph\subparagraph
\renewcommand{\subparagraph}[1]{\oldsubparagraph{#1}\mbox{}}
\fi

\title{wolensing: A Python package for computing the amplification factor for gravitational waves with wave-optics effects}

        \author[1]{Simon M.C. Yeung}
          \author[2]{Mark H.Y. Cheung}
          \author[3]{Miguel Zumalacarregui}
          \author[4]{Otto A. Hannuksela}
    
      \affil[1]{University of Wisconsin-Milwaukee, Milwaukee, WI 53201,
USA}
      \affil[2]{William H. Miller III Department of Physics and
Astronomy, Johns Hopkins University, 3400 North Charles Street,
Baltimore, Maryland, 21218, USA}
      \affil[3]{Max Planck Institute for Gravitational Physics (Albert
Einstein Institute), Am Muhlenberg 1, D-14476 Potsdam, Germany}
      \affil[4]{Department of Physics, The Chinese University of Hong
Kong, Shatin, New Territories, Hong Kong}
  \date{\vspace{-5ex}}

\begin{document}
\maketitle

\marginpar{
  \sffamily\small

  {\bfseries DOI:} \href{https://doi.org/}{\color{linky}{}}

  \vspace{2mm}

  {\bfseries Software}
  \begin{itemize}
    \setlength\itemsep{0em}
    \item \href{https://github.com/openjournals/joss-reviews/issues/}{\color{linky}{Review}} \ExternalLink
    \item \href{https://github.com/manchunyeung/wolensing}{\color{linky}{Repository}} \ExternalLink
    \item \href{}{\color{linky}{Archive}} \ExternalLink
  \end{itemize}

  \vspace{2mm}

  {\bfseries Submitted:} \\
  {\bfseries Published:} 

  \vspace{2mm}
  {\bfseries License}\\
  Authors of papers retain copyright and release the work under a Creative Commons Attribution 4.0 International License (\href{https://creativecommons.org/licenses/by/4.0/}{\color{linky}{CC BY 4.0}}).
}

\hypertarget{summary}{%
\section{Summary}\label{summary}}

The \textsc{wolensing} Python package offers a solution for
gravitational wave lensing computations within the full wave-optics
regime. This tool is primarily designed to calculate the gravitational
lensing amplification factor including diffractive effects, an essential
component for generating accurate lensed gravitational wave waveforms.
These waveforms are integral to astrophysical and cosmological studies
related to gravitational-wave lensing.

Integrating with \textsc{lensingGW} \parencite{Pagano_2020},
\textsc{wolensing} provides solutions for image positions in the
high-frequency regime where wave and geometrical optics converge. This
functionality allows the amplification factor to be applicable across a
wider frequency range. Another key feature of \textsc{wolensing} is its
ability to plot time delay contours on the lens plane, offering
researchers a visual tool to better understand the relationship between
the lens system and the amplification factor.

\textsc{wolensing} is compatible with various lens models in
\textsc{lenstronomy} \parencite{Birrer_2021}. There are also built-in lens
models including point mass, singular isothermal sphere (SIS), and
nonsingular isothermal ellipsoid (NIE) with \textsc{jax} \parencite{jax_2018} supporting GPU computation. Users can accommodate different
lens models in the code with \textsc{jax}.

\textsc{wolensing} is available as an open-source package on
\textsc{PyPI} and can be installed via \textsc{pip}.

\hypertarget{statement-of-need}{%
\section{Statement of need}\label{statement-of-need}}

Gravitational wave lensing studies have traditionally concentrated on
the strong lensing case, utilizing the geometrical optics approximation.
This approach predicts images with varying time delays and
magnifications while maintaining uniform frequency evolution across
these images. However, for lens masses around or below 100 solar masses,
the scale of the gravitational (Schwarzschild) radius becomes comparable
to the wavelength of the gravitational wave that is within the
LIGO/Virgo/Kagra sensitivity range (\(10\)--\(10^3\) Hz). Example lenses
that exist within the mass range include stars and low-mass compact
object remnants. In the LISA sensitivty range (mHz--Hz), similar
considerations apply to lenses with masses ranging from \(10^5\) to
\(10^8\) solar masses. Example lenses that exist in this mass range
include dark matter subhalos and larger compact structures. In this
limit when the Schwarzschild radius is around or below the
gravitational-wave wavelength, wave optics effects introduces
diffraction effects, necessitating a shift from geometrical to wave
optics for accurate modeling. Indeed, in this regime, the frequency
evolution of gravitational waves is influenced by the amplification
factor, which is determined using a diffraction integral, resulting in a
marked increase in computational complexity and cost.

The \textsc{wolensing} package addresses this challenge by providing
efficient computation of the amplification factor for general lenses. To
optimize computational speed, it includes built-in simple lens models
that leverage \textsc{jax} for enhanced performance. Furthermore,
\textsc{wolensing} integrates geometrical optics for high-frequency
scenarios, reducing the computational cost in that regime.

The core component of \textsc{wolensing} is a 2-dimensional integrator
that estimates the area between neighboring contour lines of the lensing
time delay function \parencite{Diego_2019}. The integration method
implemented works well for general lens systems and fine tuning of the
settings is not required when changing the lens model. Other than
scenarios with a single lens, \textsc{wolensing} can also be used to
study systems with multiple lenses. Notably, \parencite{Cheung_2021} and
\parencite{Yeung_2023} employed the package to analyze microlensing effects
on top of type-I and type-II images produced by a Singular Isothermal
Sphere (SIS) galaxy.

Other packages that could be used to compute wave-optics effects include 
\textsc{Glworia} \parencite{glworia}, which utilises contour integration to compute the amplification factor for symmetric lenses, and
\textsc{GLoW} \parencite{glow}, which employs both contour and grid integration methods to compute the amplification factor for general lenses and arbitrary impact parameters. 
The philosophy of \textsc{wolensing} is ease of usage: it is compatible with \textsc{lenstronomy}, which allows access to different varieties of lens models, making if easy to construct different lensing scenarios flexibly.

\hypertarget{acknowledgements}{%
\section{Acknowledgements}\label{acknowledgements}}

We thank Astha for useful discussion.

\hypertarget{refs}{%
\section{References}\label{refs}}

\leavevmode\hypertarget{ref-Birrer_2021}{}%
Birrer, Simon, Anowar Shajib, Daniel Gilman, Aymeric Galan, Jelle
Aalbers, Martin Millon, Robert Morgan, et al. 2021. ``Lenstronomy Ii: A
Gravitational Lensing Software Ecosystem.'' \emph{Journal of Open Source
Software} 6 (62): 3283. \url{https://doi.org/10.21105/joss.03283}.

\leavevmode\hypertarget{ref-jax_2018}{}%
Bradbury, James, Roy Frostig, Peter Hawkins, Matthew James Johnson,
Chris Leary, Dougal Maclaurin, George Necula, et al. 2018. \emph{JAX:
Composable Transformations of Python+NumPy Programs} (version 0.3.13).
\url{http://github.com/google/jax}.

\leavevmode\hypertarget{ref-glworia}{}%
Cheung, Mark Ho-Yeuk, Ken K. Y. Ng, Miguel Zumalacárregui, and Emanuele
Berti. 2024. ``Probing Minihalo Lenses with Diffracted Gravitational
Waves.'' \url{https://arxiv.org/abs/2403.13876}.

\leavevmode\hypertarget{ref-Cheung_2021}{}%
Cheung, Mark H Y, Joseph Gais, Otto A Hannuksela, and Tjonnie G F Li.
2021. ``Stellar-Mass Microlensing of Gravitational Waves.''
\emph{Monthly Notices of the Royal Astronomical Society} 503 (3):
3326--36. \url{https://doi.org/10.1093/mnras/stab579}.

\leavevmode\hypertarget{ref-Diego_2019}{}%
Diego, J. M., O. A. Hannuksela, P. L. Kelly, G. Pagano, T. Broadhurst,
K. Kim, T. G. F. Li, and G. F. Smoot. 2019. ``Observational Signatures
of Microlensing in Gravitational Waves at Ligo/Virgo Frequencies.''
\emph{Astronomy \&Amp; Astrophysics} 627 (July): A130.
\url{https://doi.org/10.1051/0004-6361/201935490}.

\leavevmode\hypertarget{ref-Pagano_2020}{}%
Pagano, G., O. A. Hannuksela, and T. G. F. Li. 2020. ``LENSINGGW: A
Python Package for Lensing of Gravitational Waves.'' \emph{Astronomy
\&Amp; Astrophysics} 643 (November): A167.
\url{https://doi.org/10.1051/0004-6361/202038730}.

\leavevmode\hypertarget{ref-glow}{}%
Villarrubia-Rojo, Hector, Stefano Savastano, Miguel Zumalacárregui, Lyla
Choi, Srashti Goyal, Liang Dai, and Giovanni Tambalo. 2024. ``GLoW:
Novel Methods for Wave-Optics Phenomena in Gravitational Lensing.''
\url{https://arxiv.org/abs/2409.04606}.

\leavevmode\hypertarget{ref-Yeung_2023}{}%
Yeung, Simon M C, Mark H Y Cheung, Eungwang Seo, Joseph A J Gais, Otto A
Hannuksela, and Tjonnie G F Li. 2023. ``Detectability of microlensed
gravitational waves.'' \emph{Monthly Notices of the Royal Astronomical
Society} 526 (2): 2230--40.
\url{https://doi.org/10.1093/mnras/stad2772}.


@article{Birrer_2021,
   title={lenstronomy II: A gravitational lensing software ecosystem},
   volume={6},
   ISSN={2475-9066},
   url={http://dx.doi.org/10.21105/joss.03283},
   DOI={10.21105/joss.03283},
   number={62},
   journal={Journal of Open Source Software},
   publisher={The Open Journal},
   author={Birrer, Simon and Shajib, Anowar and Gilman, Daniel and Galan, Aymeric and Aalbers, Jelle and Millon, Martin and Morgan, Robert and Pagano, Giulia and Park, Ji and Teodori, Luca and Tessore, Nicolas and Ueland, Madison and Van de Vyvere, Lyne and Wagner-Carena, Sebastian and Wempe, Ewoud and Yang, Lilan and Ding, Xuheng and Schmidt, Thomas and Sluse, Dominique and Zhang, Ming and Amara, Adam},
   year={2021},
   month=jun, pages={3283} }

@article{Pagano_2020,
   title={LENSINGGW: a PYTHON package for lensing of gravitational waves},
   volume={643},
   ISSN={1432-0746},
   url={http://dx.doi.org/10.1051/0004-6361/202038730},
   DOI={10.1051/0004-6361/202038730},
   journal={Astronomy &amp; Astrophysics},
   publisher={EDP Sciences},
   author={Pagano, G. and Hannuksela, O. A. and Li, T. G. F.},
   year={2020},
   month=nov, pages={A167} }

@article{Cheung_2021,
   title={Stellar-mass microlensing of gravitational waves},
   volume={503},
   ISSN={1365-2966},
   url={http://dx.doi.org/10.1093/mnras/stab579},
   DOI={10.1093/mnras/stab579},
   number={3},
   journal={Monthly Notices of the Royal Astronomical Society},
   publisher={Oxford University Press (OUP)},
   author={Cheung, Mark H Y and Gais, Joseph and Hannuksela, Otto A and Li, Tjonnie G F},
   year={2021},
   month=feb, pages={3326–3336} }

@article{Yeung_2023,
    author = {Yeung, Simon M C and Cheung, Mark H Y and Seo, Eungwang and Gais, Joseph A J and Hannuksela, Otto A and Li, Tjonnie G F},
    title = "{Detectability of microlensed gravitational waves}",
    journal = {Monthly Notices of the Royal Astronomical Society},
    volume = {526},
    number = {2},
    pages = {2230-2240},
    year = {2023},
    month = {09},
    issn = {0035-8711},
    doi = {10.1093/mnras/stad2772},
    url = {https://doi.org/10.1093/mnras/stad2772},
    eprint = {https://academic.oup.com/mnras/article-pdf/526/2/2230/51858949/stad2772.pdf},
}

@article{Diego_2019,
   title={Observational signatures of microlensing in gravitational waves at LIGO/Virgo frequencies},
   volume={627},
   ISSN={1432-0746},
   url={http://dx.doi.org/10.1051/0004-6361/201935490},
   DOI={10.1051/0004-6361/201935490},
   journal={Astronomy &amp; Astrophysics},
   publisher={EDP Sciences},
   author={Diego, J. M. and Hannuksela, O. A. and Kelly, P. L. and Pagano, G. and Broadhurst, T. and Kim, K. and Li, T. G. F. and Smoot, G. F.},
   year={2019},
   month=jul, pages={A130} }

@software{jax_2018,
  author = {James Bradbury and Roy Frostig and Peter Hawkins and Matthew James Johnson and Chris Leary and Dougal Maclaurin and George Necula and Adam Paszke and Jake Vander{P}las and Skye Wanderman-{M}ilne and Qiao Zhang},
  title = {{JAX}: composable transformations of {P}ython+{N}um{P}y programs},
  url = {http://github.com/google/jax},
  version = {0.3.13},
  year = {2018},
}

@misc{glworia,
      title={Probing minihalo lenses with diffracted gravitational waves}, 
      author={Mark Ho-Yeuk Cheung and Ken K. Y. Ng and Miguel Zumalacárregui and Emanuele Berti},
      year={2024},
      eprint={2403.13876},
      archivePrefix={arXiv},
      primaryClass={gr-qc},
      url={https://arxiv.org/abs/2403.13876}, 
}

@misc{glow,
      title={GLoW: novel methods for wave-optics phenomena in gravitational lensing}, 
      author={Hector Villarrubia-Rojo and Stefano Savastano and Miguel Zumalacárregui and Lyla Choi and Srashti Goyal and Liang Dai and Giovanni Tambalo},
      year={2024},
      eprint={2409.04606},
      archivePrefix={arXiv},
      primaryClass={gr-qc},
      url={https://arxiv.org/abs/2409.04606}, 
}
\end{document}